\documentclass[useAMS,usenatbib]{mn2e} 
\def\aap{\textit{Astron. Astrophys.}} 
\def\aj{\textit{Astron.~J.}} 
\def\apj{\textit{Astrophys.~J.}} 
\def\apjl{\textit{Astrophys.~J.~Lett.}} 
\def\apjs{\textit{Astrophys.~J.~Suppl.}} 
\def\jcap{\textit{J. Cosmol. Astropart. Phys.}} 
\def\mnras{\textit{Mon. Not. R. Astron. Soc.}} 
\def\nat{\textit{Nature}} 
\def\prd{\textit{Phys. Rev. D}} 
\def\prl{\textit{Phys. Rev. Lett.}} 
\def\prl{\textit{PASJ}}
\usepackage{dsfont}
\usepackage{amsmath}
\usepackage{amssymb}
\usepackage{graphicx}
\usepackage{verbatim}
\usepackage{natbib}
\usepackage{eucal}
\usepackage{calligra}
\usepackage{soul}

\usepackage{amsfonts}
\usepackage{color}
\usepackage[normalem]{ulem}
\usepackage[T1]{fontenc}

\usepackage{times,epstopdf}
\usepackage{epsfig}
\usepackage[usenames,dvipsnames]{xcolor}
\usepackage{url}
\usepackage{subfigure}
\usepackage{soul,xcolor}
\setstcolor{red}

\def\aj{{AJ}}                   
\def\apj{{ApJ}}                 
\def\apjl{{ApJ}}                
\def\apjs{{ApJS}}               
\def\aap{{A\&A}}                
\def\mnras{{MNRAS}}             
\def\prd{{Phys.~Rev.~D}}        
\def\prl{{Phys.~Rev.~Lett.}}    
\def\nat{{Nature}}              

\begin{document} 


\title[Dynamics of self-gravitating systems]{Dynamical analysis of stacked samples of asymmetric, non-static, self-gravitating systems}
\author[Cai et al.]
{Yan-Chuan Cai$^{1}$\thanks{E-Mail: cai@roe.ac.uk}, Nick Kaiser$^{3}$\thanks{Nick Kaiser passed away before the submission of this work.}, Shaun Cole$^{2}$ and Carlos Frenk$^{2}$ \\
$^{1}$ Institute for Astronomy, University of Edinburgh, Royal Observatory, Blackford Hill, Edinburgh, EH9 3HJ , UK \\
$^{2}$ Institute for Computational Cosmology, Department of Physics, Durham University, South Road, Durham DH1 3LE, UK \\
$^{3}$D\'epartement de Physique, \'Ecole Normale Sup\'erieure, Rue d'Ulm, Paris  75005, France
}
\maketitle  

\begin{abstract}
We use numerical simulations to explore biases that arise in dynamical estimates of the mean mass profile for a collection of galaxy clusters that have been stacked to make a composite. There are three types of bias. One arises from anisotropy of the kinematic pressure tensor and has been already well studied; a second arises from departures from equilibrium; and a third arises because of heterogeneity of the clusters used, from their individual non-sphericity, and because velocities used are measured with respect to centres that are, in general, accelerating. Here we focus on the latter two. We stack clusters to measure the pressure tensor and density profiles and then estimate the dynamical mass profile using the Jeans equation, and compare to the actual mean mass profile. The main result of this paper is an estimate of the bias, that can be used to correct the dynamical mass estimate, and we show how it depends on the cluster sample selection. We find that Jeans equation typically overestimates the true mass by about 20\% at the virial radius.

\end{abstract}


\maketitle

\section{Introduction}

Ensembles of clusters `stacked' in redshift space have proved to
be very useful in cosmology.
A pioneering study was carried out by the CNOC collaboration (Canadian Network for Observational Cosmology) in
the '90s and similar studies were pursued subsequently \citep{Carlberg1997B,Carlberg1997C,  Carlberg1997A,vanderMarel2000}. The goal of these studies was to use the
virial theorem or the Jeans equation to obtain the mass-to-light
ratio and thereby, on multiplying by the luminosity density, an estimate of the density parameter (subject to assumptions about
biasing of galaxies with respect to the mass).

The benefits of carrying out this type of analysis using
stacked rather than using individual clusters may
seem to be self-evident. Individual clusters often appear to be irregular and, in the context of hierarchical clustering, are not
expected to be well `relaxed' or equilibrated.
A stacked cluster, in contrast, will be highly symmetric and, if constructed
from large numbers of clusters, will be evolving on a cosmological time 
frame -- i.e.\ very slowly -- and would be expected to be
essentially in equilibrium within the virial radius and
subject to only  smooth symmetrical infall and with a slow rate of change
of its momentum density outside the virial radius.  In short, it
would seem that a stacked sample would obey, to a very
good approximation, the assumptions that are made in applying
the Jeans equation to relaxed stellar systems such as globular clusters
or elliptical galaxies.

The CNOC study -- large for its time -- targeted a dozen clusters. 
There are now catalogues of hundreds of thousands of clusters that
have been extracted from the SDSS survey \citep[e.g.][]{Yang2007, Hao2010, Wen2012,Tempel2012, Rykoff2014, Lim2017} though these catalogues
extend down to lower masses than those of the CNOC clusters.

This has enabled a measurement of the gravitational redshift in
clusters \citep{Wojtak2011, Sadeh2015, Jimeno2015, Alam2017, Mpetha2021, Rosselli2023}, a {\em tour de force\/} since the effect is
of second order in the velocity dispersion and therefore requires
very large numbers of redshifts to beat down the `root-N' noise from
the first order Doppler effect.

The future prospects are still brighter. 
Several projects are under way to generate large samples of clusters and at higher
redshifts. Massive spectroscopic ground-based surveys such as DESI \citep{DESI, DESIHahn2023} are already delivering data. Over one million groups and clusters have been derived from the DESI Legacy Imaging Survey data \citep{Yang2021,Wen2024}. Satellite missions such as Euclid and the Nancy Grace Roman Space Telescope \citep{Euclid,WFIRST} promises to deliever deep samples of clusters. Tens of thousands of optically confirmed galaxy groups and clusters have also been detected in X-ray by the SRG/eROSITA All-Sky Survey \citep{Bulbul2024}.

One appealing cosmological test is to use a stacked sample of a large number of massive clusters to generate a stacked projected mass density profile. This can then
be compared to the mass density profile inferred from the weak lensing
of background galaxies that will be measured in deeper photometry. 

\begin{figure*}
\begin{center}
\scalebox{0.48}{
\hspace{-1.3 cm}
\includegraphics[angle=0]{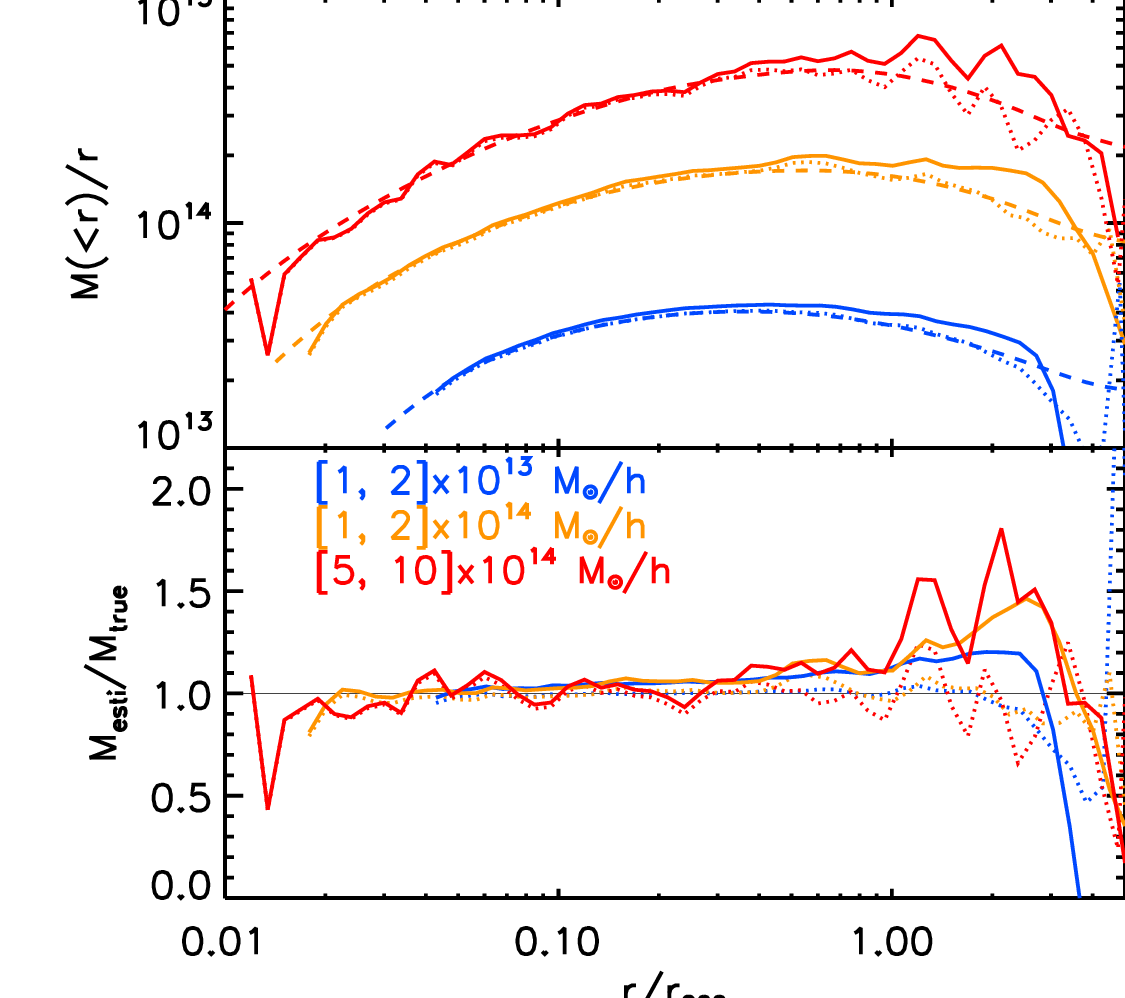}}
\vspace{0.4 cm}
\scalebox{0.48}{
\hspace{-3.7 cm}
\includegraphics[angle=0]{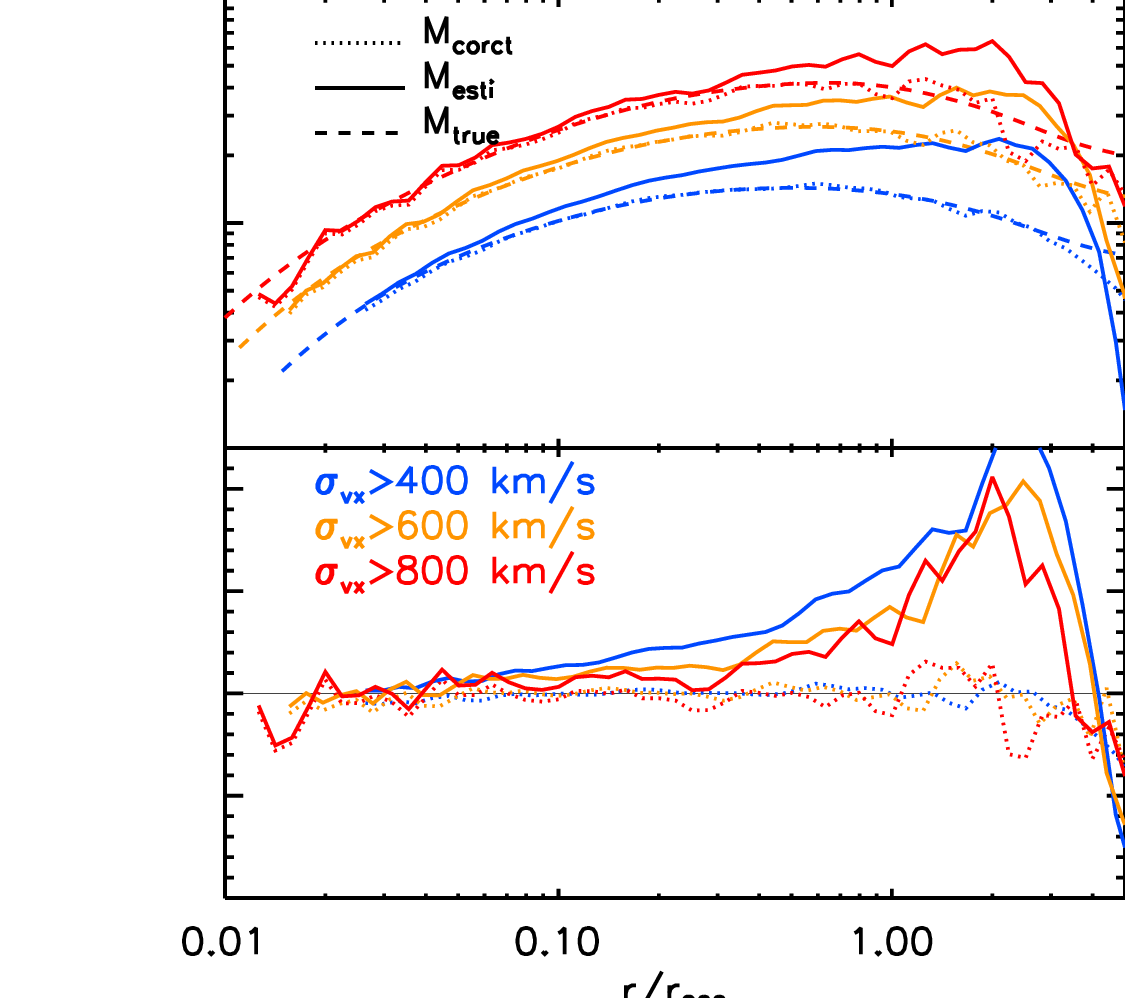}
}
\vspace{-0.3 cm}
\caption{The different coloured lines show masses estimated using dynamical analysis of a stack composed of clusters of different masses and velocity dispersions as a function of radius. The radius is expressed in units of $r_{200}$, the radius interior to which the density is $200$ times the critical density as calculated from the mean mass of each stacked cluster sample.
They are 0.4, 0.8 and 1.4 $h^{-1}$Mpc (0.8, 1.0 and 1.3 $h^{-1}$Mpc) for the three mass ranges (velocity dispersion ranges) respectively. The most bound particles of the most massive sub-haloes in each FoF group are adopted as proxies of BCGs which are taken as the halo centres. BCG velocities are taken to be the average of all particles within a radius $r_{\rm c}=30h^{-1}$kpc.
On the left are shown stacks made from samples of
limited mass range while on the right are shown `cumulative' stacks made using
all clusters with velocity dispersion above the limiting value indicated in the legend.
In each case we have measured the mean kinetic pressure gradient tensor and the density
of haloes and combined this according to the Jeans equation to obtain an estimated
mass. For a single spherically symmetric cluster with the same pressure and
number density profile this would give the true mass. But for a composite this
is not the case.  The correction factors plotted here can be used to correct
estimated mass profiles to obtain the true mean mass profile which can then be
used to predict e.g.\ the mean weak lensing signal for background galaxies.
Top panels:  Solid lines show the estimated mass profiles $M_{\rm esti}$ from the kinematic pressure gradient given on the right-hand side of (\ref{Eq:NaiveM}), which deviate from the true mass profiles (dashed lines) at
large radii. Dotted curves shows the estimated mass profiles {$M_{\rm corct}$} after applying the correction terms shown in (\ref{Eq:F1}) \& (\ref{Eq:F2}), i.e. $F_1F_2M_{\rm esti}$. Bottom panels: the ratio between the estimated mass, with (dotted) or without (solid) corrections  versus the true mass. 
 }
\label{Fig:MassProfile}
\end{center}
\end{figure*}

There are several interesting ways this comparison can fail to give the
expected results. One is the assumption that the galaxies in the clusters are
following geodesics. Modified theories
of gravity predict that galaxies, when unscreened, would be subject to additional
`fifth forces' \citep[e.g.][]{Zhang2007, Schmidt2010, Zhao2011, Lombriser2012, Zu2014,Arnold2014, Falck2015, Gronke2015, Pizzuti2024,Butt2025}. Second, as the 
test involves comparison with the lensing signal there is the additional
assumption that the photons follow geodesics and are thus sensitive to
modifications that change the standard GR relation between the temporal and
spatial parts of the metric perturbations.  
Third, and essentially unrelated to the physics in the cluster, 
the lensing derived mass also depends, in a geometric manner
on the distance-redshift relation.  This provides a useful
cross-check on the validity of the cosmological model inferred from
other probes (CMB anisotropies and the distance-redshift relation in
the more local universe) \citep[e.g.][]{Jain2003, Biesiada2010}.

To be effective, such tests need to be both precise and accurate.
Precision is a given; current surveys provide very large numbers of
clusters and redshifts.   
Accuracy is more challenging. One well known problem is the
difficulty of determining the velocity dispersion anisotropy, 
characterised by $\beta$ \citep[e.g.][]{Richardson2013,Read2017,Genina2020}. The more radial the anisotropy, the lower the mass that is inferred
from the Jeans equation (which uses only the second moment of the
velocity dispersion).  
This problem can be addressed in two ways; using N-body simulations
to calibrate the effect of the velocity dispersion anisotropy, and using higher 
than second order moments of the velocity distribution to constrain $\beta$.

Here we will assume that the anisotropy problem is solved to
adequate accuracy and that, from measurements of the distribution
of redshifts as a function of projected separation and knowledge of
$\beta$ we have at our disposal the ensemble average pressure tensor
$\overline{n \langle v_i v_j \rangle}$ as a function of radius from the cluster
centre. The overbar represents ensemble averaging 
over the cluster sample. For an equilibrated individual cluster
this would be sufficient to determine the mass profile. 
This is not the case for a stacked cluster - even though it
may be essentially perfectly symmetrical and equilibrated. In this paper we explore this effect in numerical simulations. Our goal is to make a first attempt at
quantifying the resulting biases in order to be able to obtain
from stacked clusters from future surveys accurate predictions
for e.g.\ the mean gravitational lensing signal.

\section{Halo mass estimate from the Jeans equation:} 

The Jeans equation is the first moment of the collisionless Boltzmann equation
\begin{equation}
\label{Eq:Jeans1}
\partial_t n\langle v_i\rangle +\partial_j(n\langle v_iv_j\rangle)+n\partial_i\Phi=0,
\end{equation}
where for simplicity, we have dropped the time and spatial dependence for all the quantities. Here
$n({\bf r}, t)=\int f({\bf r, v}, t)d^3v$
is the density of the `tracer particles', 
$\langle v_i\rangle = n^{-1}\int f({\bf r, v}, t)v_id^3v$ is their mean velocity
and $\langle v_i v_j\rangle= n^{-1}\int f({\bf r, v}, t)v_iv_jd^3v$ their velocity dispersion tensor. Here $f({\bf r, v}, t)$ is the phase space density (PSD) of 
tracer particles (galaxies in the case of a cluster)\footnote{In the context of this paper, we consider galaxies as collisionless particles like cold dark matter, and use the terms galaxy and dark matter particles interchangeably, ignoring the possible biases between them.},
and $\Phi$ is the Newtonian gravitational potential, conventionally defined
with respect to its value at spatial infinity.

The Jeans equation expresses the conservation of momentum 
for a population of tracer particles moving in the potential $\Phi$.
The essential physical assumption on which it is based is that
the tracers are unable to exchange momentum with other constituents
such as other tracers or the dark matter.
Equation (\ref{Eq:Jeans1}) states that
the rate of change of the momentum density in a volume element 
is the divergence of the
flux of momentum -- the rate at which pressure gradients are delivering
momentum, or the force density -- minus the rate at which gravity is
removing momentum.  

The Jeans equation is sometimes solved to provide the phase-space density
for an assumed potential.  Here we think of the PSD as given and wish to
use it to obtain the potential. Note that this equation can be applied to any tracer
population and should provide identical results for the potential. This was confirmed for the case of red and blue galaxies in the CNOC sample by \citep{Carlberg1997A}.

For a static equilibrated system the first term vanishes.  For a {\em spherical\/}
system the gradients of the kinetic pressure and the gravity are
both radially directed and Gauss's law provides the mass profile of a cluster 
\begin{equation}
M(<r)=- r^2 \tilde g(r)/G ,
\label{Eq:NaiveM}
\end{equation}
{where $\tilde g \equiv {\hat r}_i\partial_j( n\langle v_i v_j\rangle) /n$
is the gravity inferred from the kinetic pressure gradient.}
The mass here includes the negative effective mass $M_\Lambda = - \Omega_\Lambda H^2 r^3 / G$ of the cosmological constant or quintessence field.  The actual mass that
generates the weak lensing signal, for instance, is $M_{\rm matter} = M(<r) - M_\Lambda$.

As discussed, a stacked cluster sample is expected to be both highly 
spherical and equilibrated (at least at small radii).  
It therefore might seem reasonable to
assume that one can use (\ref{Eq:Jeans1}) applied to the stacked sample
in the manner discussed here
to obtain the ensemble average mass profile as in (\ref{Eq:NaiveM}), with perhaps
a correction for the non-vanishing rate of change of momentum
density in the infall region \citep[e.g.][]{Cuesta2008}. However, we show in Fig.~\ref{Fig:MassProfile} that this is not true, 
i.e. the mass of the stacked cluster inferred from the pressure 
gradient is usually higher than the true mass. Especially around  
1-2 times $r_{\rm 200}$ where the estimated mass can be a factor of 2 too high. Why? As we shall now describe, things are
a little more complicated and the naive application of (\ref{Eq:Jeans1})
results in a biased estimate of the mass -- even if the pressure
gradient is perfectly known.

While appealingly simple, there are serious questions of principle 
regarding application of (\ref{Eq:Jeans1}) 
to a real composite. {One of these stems} from the fact that equation (\ref{Eq:Jeans1})
is a Newtonian equation and is therefore valid in Newton's absolute
space; one of a family of rigid frames that are not accelerating with respect
to one another.

For a single isolated cluster such a frame could be realised as 
a light but rigid lattice anchored to masses at larger distance, and
the Newtonian velocities are those that would be measured locally
by observers tied to the lattice.
The analogue in cosmology would be velocities
with respect to observers that are fixed in conformal coordinates
(plus the uniform expansion velocity of such a frame).
But neither case is realistic as these fictitious
observers are being accelerated by the lattice.
In practice what we have access to are velocities
relative to some centre in the cluster -- perhaps the brightest cluster
galaxy (BCG) -- and that reference frame is in free
fall. One might imagine applying equation (\ref{Eq:Jeans1}) with such observed relative velocities, but that
would mean, for instance, that from local observations within a
gravitating system one can determine $\partial_i \Phi$, but that
would be nonsensical as $\partial_i \Phi$ contains a part due to
the attraction or repulsion of distant structures.

The resolution is straightforward: (\ref{Eq:Jeans1}) is not valid for observed velocities; rather,
if the velocities are measured with respect to a freely falling
particle such as a BCG, rather than with respect to a fictitious
non-inertial Newtonian observer, then the generalisation
of the Jeans equation is
\begin{equation}
\partial_t n\langle v_i\rangle+\partial_j(n\langle v_iv_j\rangle)+n a_{0i} + n\partial_i\Phi=0,
\label{Eq:a0}
\end{equation}
where ${\bf a}_0$ is the gravitational acceleration of the centre. 

Equation (\ref{Eq:a0}) tells us that local measurements within a gravitating system
can only tell us about ${\bf a}_0 + {\boldsymbol \nabla}\Phi$.  This is reassuring
as it means that the effect of distant structures cancels out.
In dynamical studies of individual galaxies or clusters, the gravitational acceleration of the centre is usually ignored \citep[e.g.][]{Cappellari2008, Watkins2013, Mamon13, Old2014, Shi2024}. This is justified because usually what one is interested in is
not the potential gradient ${\boldsymbol \nabla}\Phi$ itself but rather
the Laplacian of the potential, which of course provides the density
via Poisson's equation. In that case it does not matter
whether one dots ${\boldsymbol \nabla}$ with ${\boldsymbol \nabla}\Phi$ or ${\boldsymbol \nabla}\Phi + {\bf a}_0$ 
since ${\bf a}_0$ has no dependence on position.

But if one is trying to do a dynamical analysis of a {\rm stacked cluster\/} then it
is important to take the acceleration of the centre into account since
the acceleration of the centre is determined in part by, and therefore
correlates with, the density at the location where one is using
(\ref{Eq:a0}) to determine the gravity.

In addition to peculiar accelerations of centres, dealing with realistic cluster centres -- such as using BCGs -- might appear, at first sight,
to present formidable challenges.  A crude approximation might be to model the clusters
as individually spherical -- with say NFW \citep{NFW} -- profiles and to model the BCGs as being
drawn from a relatively cold population.  Observations suggest that the RMS velocity
of the BCG relative to the centre of mass (COM) is perhaps $\sim 40\%$ of the total
velocity dispersion \citep[e.g.][]{Becker2007, Lauer2014}.  The finite motion of the BCGs will inflate the velocity dispersion,
so one can correct for this by a suitable subtraction in quadrature (i.e.\ dividing the
observed velocity dispersion by $1 + 0.4^2$ in this case).  This was the approach taken by \citep{Wojtak2011}.  But if one were going to do that one would
also want to allow for the fact that if the BCGs are not perfectly cold then they
will also generally be displaced from the true cluster centre \citep[e.g.][]{Harvey2017,DePropris2021}. This will bias the observed density profile $n(r)$, so one might feel inclined to try to correct
for this to obtain the relevant data to use in the Jeans equation. And that is under
the, not-very-good, assumption that clusters are individually spherical.

These subtleties in applying Jeans equation to a composite are not the whole story.
Fig.~1 also shows that there is a significant difference between the bias for
a stack composed of clusters with a limited mass range (left hand plots) and for
a `cumulative' samples of clusters comprising all of the clusters above
some lower limit on mass or velocity dispersion (right hand plots). This is
not altogether surprising. As all the terms are linear, it is legitimate to average the terms in (\ref{Eq:a0}), to obtain 
\begin{equation}
\overline{n g_i}-\overline{na_{0i}}=
\partial_t \overline{n\langle v_i\rangle}+\partial_j(\overline{n\langle v_iv_j\rangle}),
\label{Eq:stacked}
\end{equation}
where $g_i=-\partial_i \phi$.
However, for
the average mass within a sphere one needs the
average gravity integrated over the surface of the  sphere, 
\begin{equation}
    M(<r) = r^2 \overline{\int\hat{r_i}(g_i-a_{0i}) d\Omega}/G,
\end{equation}
not the gravity weighted by the tracer density 
$\overline{n g_i}$  that appears in (\ref{Eq:stacked}).

Nevertheless we can manipulate (\ref{Eq:stacked})
to re-express this equation 
for the average enclosed mass of the stacked cluster and produce a corrected version of
(\ref{Eq:NaiveM}):
\begin{equation}\label{Eq:mass}
M(<r)=-r^2 \tilde{g}(r) F_1(r)F_2(r)/G,
\end{equation}
where $\tilde{g}(r) \equiv \hat{r}_i \partial_j \overline{n\langle v_i v_i \rangle}/\overline{n}$ is the radial pressure gradient divided by the mean density. This is identical to (\ref{Eq:NaiveM}) but with two correction factors (both of
which are functions of radius).
The first is
\begin{equation}
F_1(r)=1+\frac{\hat r_i \partial_t\overline{n\langle v_i\rangle}}{\hat r_i \partial_j \overline{n\langle v_iv_j\rangle}},
\label{Eq:F1}
\end{equation}
and it corrects for the neglect of the 
term 
that describes
rate of change of momentum density, $\partial_t \overline{n \langle v_i \rangle.}$
The second is
\begin{equation}
F_2(r)=\frac{\overline{\int { \hat r_i [g_i({\bf r})-a_{0i}]} d\Omega}} { \bar n(r)^{-1}\overline{\int { \hat r_i [g_i({\bf r})-a_{0i}] n({\bf r})}d\Omega}},
\label{Eq:F2}
\end{equation} 
which incorporates the complications that dynamical analysis tells us the
difference between the gravitational acceleration and the acceleration of the
centre and the fact that the ensemble average of the first term in (\ref{Eq:stacked}) is
weighted by the density of galaxies, while Gauss's law relates the mass interior
to some spherical surface to the integral of $g$ averaged by solid angle over the surface. So by definition, the numerator is the same as $-M(<r)/r^2$, where $M$ is now the averaged mass of the stacked cluster.

In this paper we use the Millennium N-body simulation to estimate the correction terms 
$F_1$ and $F_2$ as defined by equations (\ref{Eq:F1}) and (\ref{Eq:F2}).  
{Our goals are {\it i}\/) determine if these effects are significant and 
{\it ii}\/) attempt to estimate their magnitude so that the naive dynamical mass
may be corrected to obtain the true mass.}

{In outline, what we have done is to identify clusters in the final output of the
simulation {as spheres of mass which are 200 times the critical density and centred on the most bound particles of the most massive sub-haloes within each friends-of-friends group \citep{Davis1985, Springel2008},} 
yielding 35300 haloes with $M>10^{13} h^{-1}$M$_{\odot}$ in the simulation box.
As a proxy for the velocity ${\bf v}_0$ and acceleration ${\bf a}_0$ of the BCGs we have averaged the peculiar velocity and the peculiar gravity for 
particles within a core radius of $30 h^{-1}$kpc from each cluster centre.
We have determined physical positions and relative velocities of the other particles with respect to this and we have stacked these particles in physical coordinates
to obtain quantities like the number density and the components of the pressure tensor.
In order to calculate the
time derivative of the momentum density -- the first term in Jeans equation --
we have advanced all the particles from the $a = 1$ output to a future
epoch when $a = 1.04$ and re-determined centres and relative positions and
velocities tracking the same core particles identified at the $a=1$ epoch.
We made two stacks and
differenced the mean radial momentum density and divided by the time
interval. In the plots below we show the results for
various sub-samples of clusters selected according to mass or velocity 
dispersion. In each case the quantities of interested are plotted as a function
of $r / r_{200}$, where $r_{200}$ is the average of the usual 200 times critical
density radius taken over the sample of clusters.  More details of the
simulations and the procedures used are given in the Appendix.}

\begin{figure*}
\begin{center}
\scalebox{0.48}{
\hspace{-1.3 cm}
\includegraphics[angle=0]{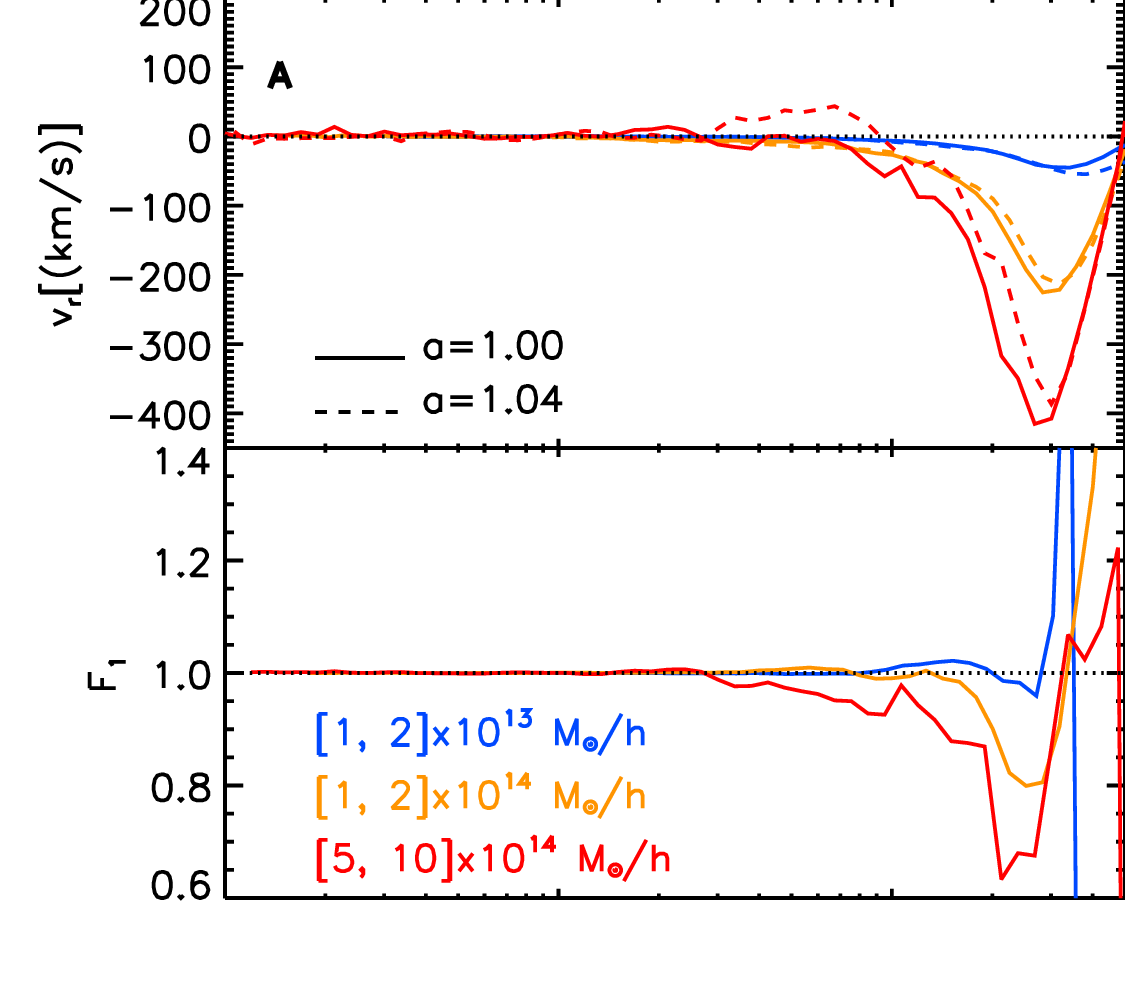}}
\vspace{-0.66cm}
\scalebox{0.48}{
\hspace{-3.7 cm}
\includegraphics[angle=0]{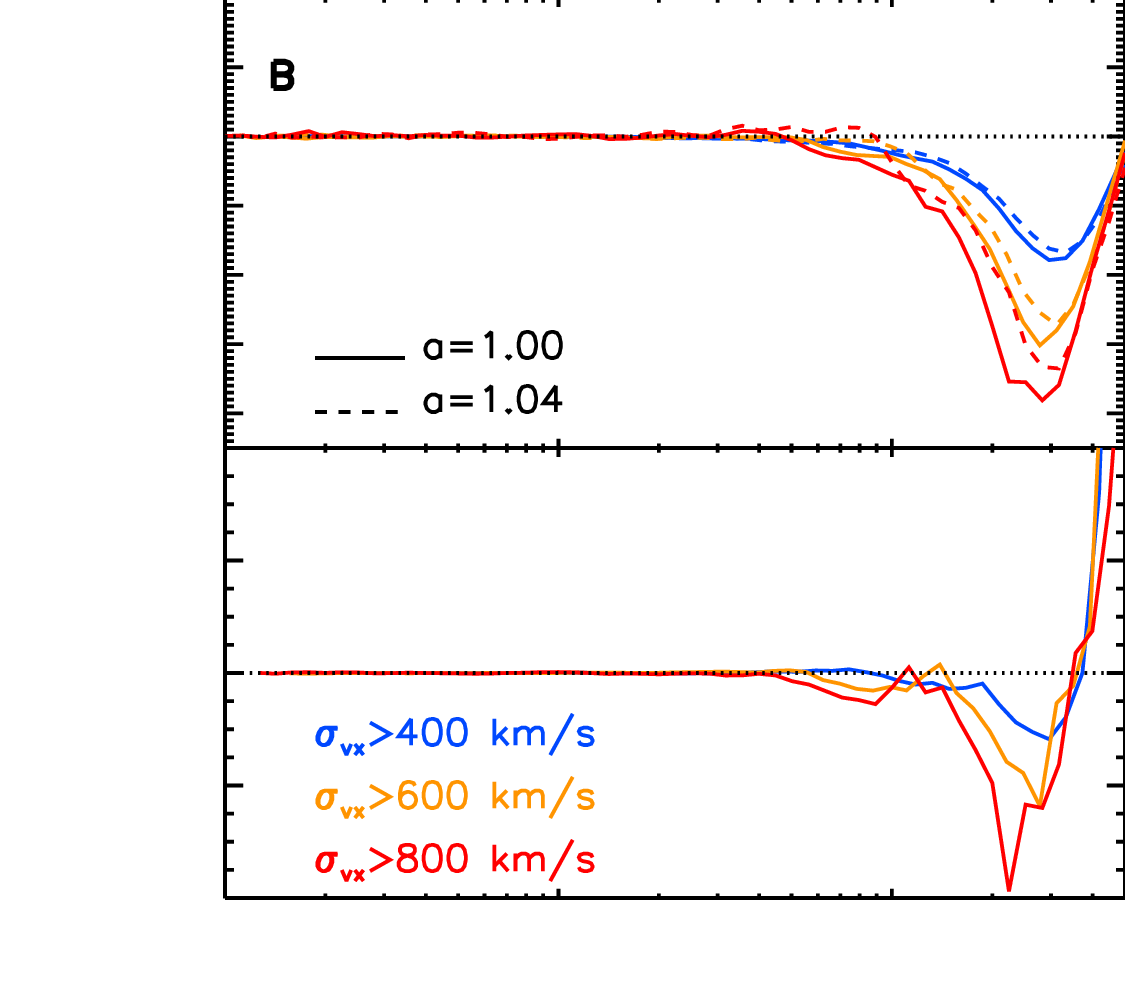}}
\scalebox{0.48}{
\hspace{-1.3 cm}
\includegraphics[angle=0]{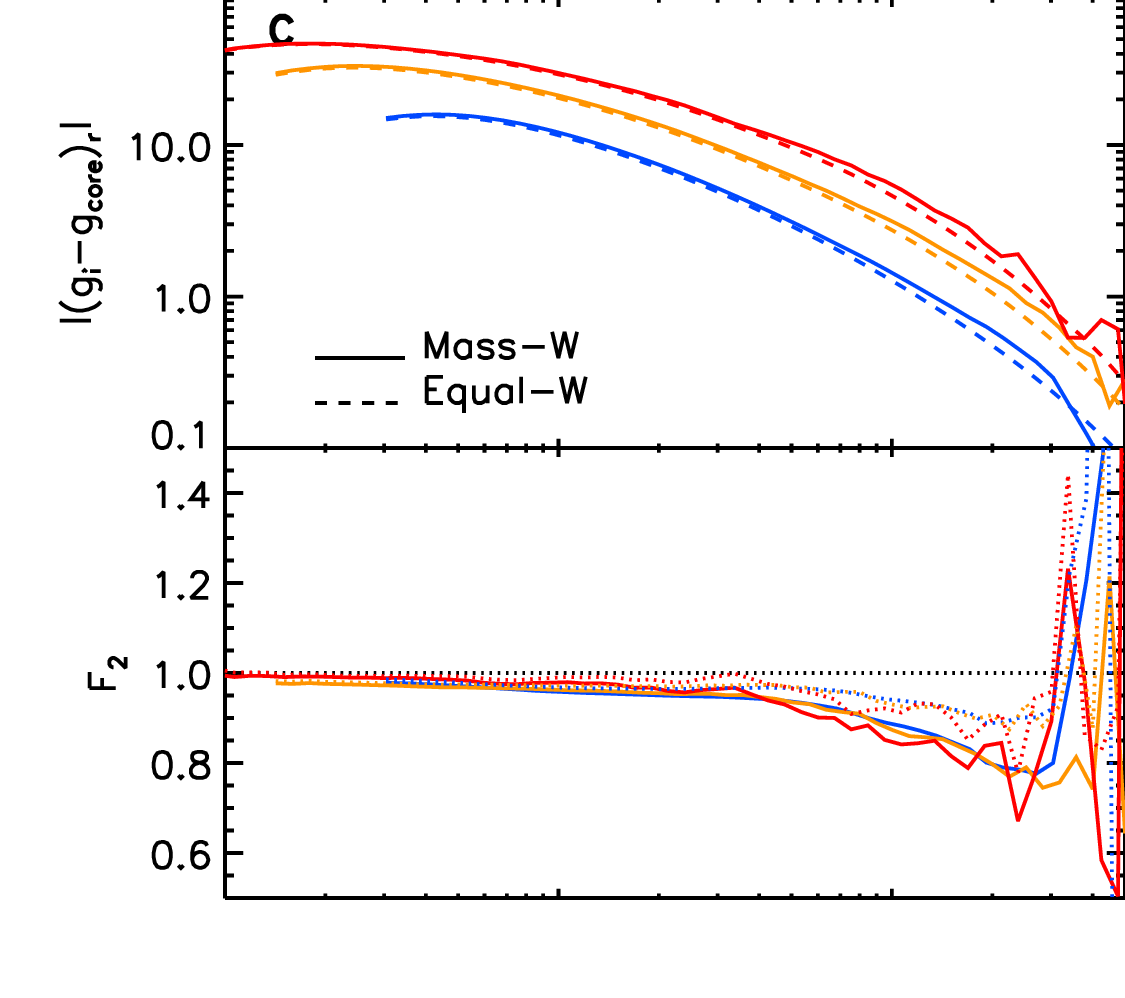}}
\vspace{-0.66cm}
\scalebox{0.48}{
\hspace{-3.7 cm}
\includegraphics[angle=0]{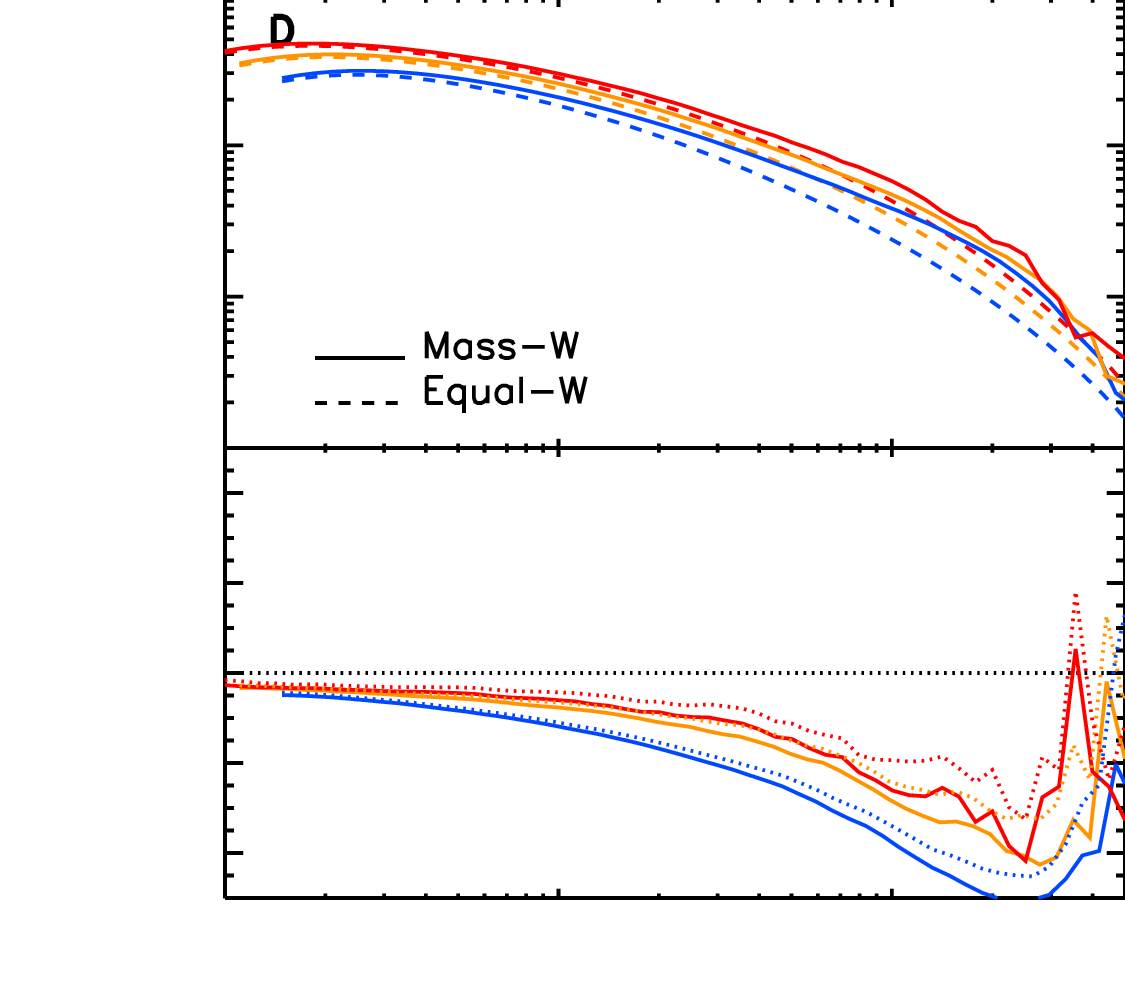}}
\scalebox{0.48}{
\hspace{-1.3 cm}
\includegraphics[angle=0]{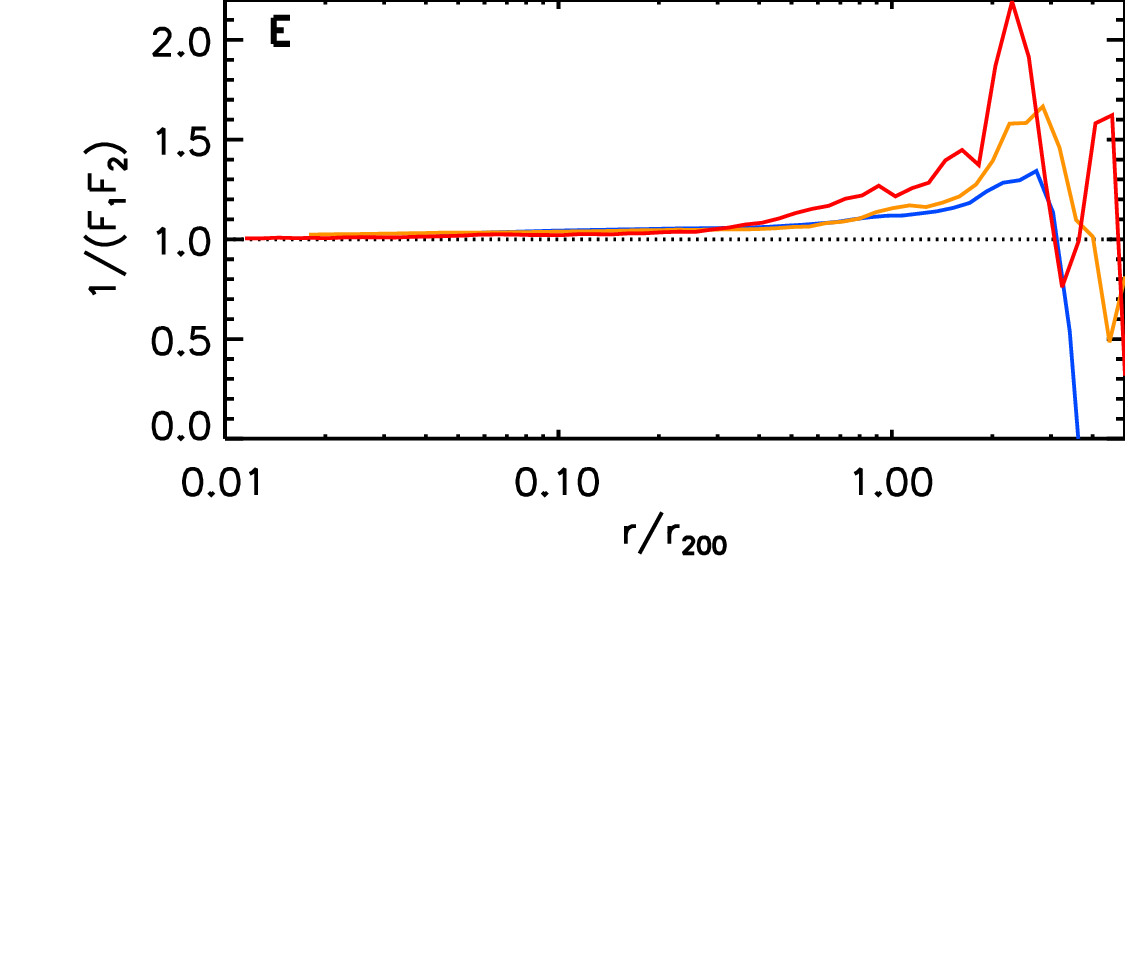}}
\scalebox{0.48}{
\hspace{-3.7 cm}
\includegraphics[angle=0]{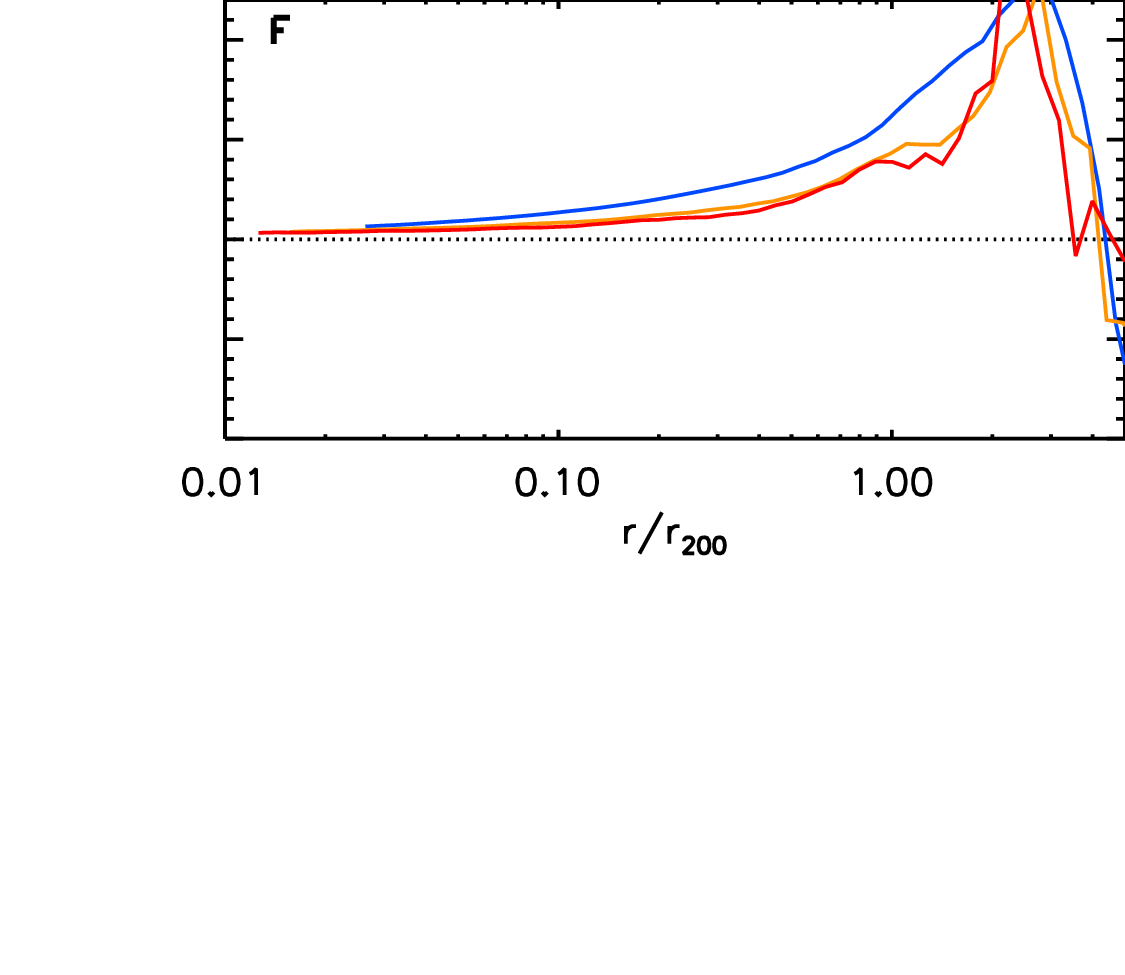}}
\vspace{-3.4 cm}
\caption{The top panels of A \& B show the mean radial velocity profiles at two different output times, from which we calculate the correction term $ F_1$, which is plotted immediately below. Left panels show the results for samples of clusters selected in limited mass ranges, and right panels show the results for clusters within cumulative velocity dispersion ranges. 
The solid and dashed lines in A \& B show the profiles at $a=1$ and $a=1.04$ respectively.  
Bottom panels of A \& B: the solid lines show the factor  $F_1$.
C \& D: The factor $F_2$ from equation (\ref{Eq:F2}) and the quantities from which it is derived.
Within C \& D, the top panels are the radial acceleration profiles. Solid and dashed lines represent 
mass weighted ({Mass-W}) [the denominator of (\ref{Eq:F2})] and equal weighted ({Equal-W}) [the numerator of (\ref{Eq:F2})] versions respectively. Bottom: {The solid lines represent the} factor $F_2$ determined from the ratio of these two weightings. {The dotted colour lines show the case for $F_2$ when setting the central acceleration ${\bf a_0}=0$}.   E \& F: The combined correction factor $F_1F_2$ from the product of $F_1$ and $F_2$. Note that $1/ (F_1F_2)$ is the total bias. These have been used to correct the estimated halo mass profiles shown in Fig.~\ref{Fig:MassProfile}.}
\label{Fig:F1}
\end{center}
\end{figure*}

\section{Quantifying the biases}

\subsection{The $F_1$ term} 

The factor $F_1$ accounts for the 
time variation of the radial momentum density, which typically becomes important 
in the outskirts of a halo, and results in an imbalance between the
divergence of the momentum flux associated with the kinetic pressure and
the rate at which gravity is removing momentum.

It is convenient to work in spherical coordinates where (\ref{Eq:F1}) 
may be written as 
\begin{equation}
F_1(r)=1+\frac{\partial_t\overline{n(r)\langle v_r(r)\rangle}}{\partial_r\overline{n(r)\langle v_r^2(r)\rangle}+2\overline{n(r)[\langle v_r^2(r)\rangle-\langle v_{\bot}^2(r)\rangle]}/r},
\label{Eq:F1_2}
\end{equation}
where the subscripts $r$ and $\bot$ indicate the radial and tangential directions. Equivalently,
and for computational convenience,
we may write
$F_1(r)=1+f_1/(f_2+f_3)$, where 
\begin{equation}
f_1  = \frac{\partial_t\overline{n(r)\langle v_r(r)\rangle}}{\overline{n(r)}}, 
\end{equation}
\begin{equation}
f_2 = \frac{\partial_r\overline{n(r)\langle v_r^2(r)\rangle}}{\overline{n(r)}}, 
\end{equation}
and
\begin{equation}
f_3 =  \frac{2\overline{n(r)[\langle v_r^2(r)\rangle-\langle v_{\bot}^2(r)\rangle]}/r}{\overline{n(r)}}. 
\end{equation}

In the uppermost panel of Fig.~\ref{Fig:F1} we show the radial velocity profiles of stacked haloes
for the case of samples with relatively narrow mass ranges on the left and
`cumulative' samples on the right where we include all clusters with velocity dispersion
exceeding the limits shown in the legend.
We see little or no infall within $r_{200}$, suggesting that these haloes have stopped accreting mass within $r_{200}$. Between 1-3 $r_{200}$, the amplitude of infall increases with halo mass. The peak of mass infall occurs around $2r_{200}$ \citep[see also][]{Cuesta2008, Fong2021, Falco2013}\footnote{The distance at which we find the infall peaks seems larger than was reported by \citep{Cuesta2008}, i.e. $\sim 2 r_{\rm vir}$. This is due to differences in halo definitions. In  \citep{Cuesta2008}, the halo radius,  $r_{\rm vir}$, is defined by the overdensity $\Delta_c=18\pi^2+82x-39x^2$ is defined, where $x=\Omega(z)-1$ \citep{BryanAndNorman1998}. $\Delta_c$ is significantly smaller than $200, \rho_{\rm crit}=200 \bar{\rho}/\Omega_m$. Therefore, $r_{200}$ adopted here is smaller than $r_{\rm vir}$ in \citep{Cuesta2008}.}. The subplot  immediately below shows the correction factor, $F_1$, defined in (\ref{Eq:F1_2}).

The factor $F_1$ deviates more from unity as halo mass increases (left hand panels). For haloes with 
$M\sim 10^{14}h^{-1}$M$_{\odot}$, $F_1$ is about 20\% lower than unity at around 2-3 $r_{200}$. For $M\sim 10^{13}h^{-1}$M$_{\odot}$, the shift is $\sim 5-10\%$. 
Similar results are found with haloes selected using their 1D LOS velocity dispersions shown on the right-hand panels of Fig.~\ref{Fig:F1}. We also find similar results when choosing the centre of mass rather than the BCG as the halo centre, as expected, because the evolution of the radial infall only occurs at the outskirts of the halo, which is far from the scale where the choice of halo centre matters.

Our analysis is similar in some ways to that of \citet{Falco2013}. They write down Jeans equation for a cluster that is assumed to be spherically symmetric but may have non-zero radial peculiar velocity. 
They then construct a stacked cluster using a N-body simulation, much as we have done here,
and make analytic fits for the quantities density, radial velocity and velocity dispersion and
anisotropy parameter $\beta$ as a function of radius.  One of their
plots (the bottom panel of their figure 3) shows the ratio of the true velocity dispersion to
that which would have been obtained for the same density and $\beta$ profile if
the radial peculiar velocity were zero.  This shows differences at the $\sim 5$\% level
at around the virial radius (i.e.\ about 10\% for $\sigma^2$).  Assuming this may be
interpreted as the level of error in the mass estimate that one would incur by neglecting net radial motions, their result seems to be broadly in line with what we have found here.

\subsection{The $F_2$ term}

The $F_2$ factor in equation (\ref{Eq:F2}) incorporates the effect of the acceleration of the cluster centre and the fact that the estimated radial acceleration is (galaxy) number weighted rather than isotropically weighted. 

The numerator is the true mass of the stacked sample divided by $r^2$. We denote this as $g^{\rm Equal-W}$. The number weighted mass appearing in the 
denominator (denoted as  $g^{\rm Mass-W}$) differs from the true mass for three reasons: a) $\bf g(r)$ is anisotropic and the anisotropy is directly related to the distribution of mass (or galaxies). This means that $g^{\rm Mass-W}$ receives more weight in regions of acceleration where there are more galaxies (mass); b) the central acceleration ${\bf a_0}$ averages to zero for $g^{\rm Equal-W}$, but this is not the case for $g^{\rm Mass-W}$; The effect of a non-zero ${\bf a_0}$ for $F_2$ is therefore correlated with the level of non-sphericity of haloes. To separate these two contributions, we will also investigate the situation where ${\bf a_0}$ is set to be zero. c) Each stacked halo contributes to $g^{\rm Equal-W}$ equally while their contribution to $g^{\rm Mass-W}$ is proportional to the number of galaxies (or mass) it hosts. 

It can be seen from subplots C \& D of Fig.~\ref{Fig:F1} that $F_2$ is not unity at all scales. It is smaller than unity within $3r_{200}$, i.e. the mass weighted radial acceleration $g^{\rm Mass-W}$ is greater than $g^{\rm Equal-W}$. In general, the deviation of $F_2$ from unity increases with radius. {This is the joint consequence of the non-zero acceleration of halo centres and the anisotropic distribution of mass within haloes.} {When setting ${\bf a_0}=0$, the deviations of $F_2$ from unity are reduced by about a factor 2 (dotted lines at the lower panel of C). This suggests that the non-sphericity of individual haloes contributes at about the same level as the effect of a non-zero ${\bf a_0}$} 
At radii greater than $3{r_{200}}$, $F_2$ rises above unity and then drops back. This is mostly due to the presence of neighbouring haloes and the anisotropic cosmic-web structures around each centre. A neighbouring halo has its own associated local potential minimum. As one approaches  the neighbour from the the main halo, the slope of the potential may reverse in  sign \citep[see also][]{Cai2016}. In this region, the local radial acceleration points outwards from the main halo centre (but inwards with respect to the centre of the neighbour). This reduces the amplitude of the mass weighted radial force more than in the case of equal weighting. Therefore, the amplitude of $g^{\rm Mass-W}$ becomes smaller than that of $g^{\rm Equal-W}$.

Quantitatively, when haloes are binned according to their mass (left hand panels of Fig.~\ref{Fig:F1}), $F_2$ is smaller than unity by $\sim 10-20\%$ at $r_{200}$ and $\sim 20-40\%$ within 2-3${r_{200}}$, which means a bias of the same amounts for halo mass estimates if these effects are neglected.
{There is no obvious dependence on halo mass bin for $F_2$.} When haloes are selected above a threshold velocity dispersion (right hand panels of Fig.~\ref{Fig:F1}), we find that the bias becomes larger when including a larger dynamical range of halo velocities (masses). For example, with $\sigma_{\rm vx}>400$km~s$^{-1}$, the bias is as large as 50\% (blue curves in the right hand panels). This can be explained by the fact that the numerator in (\ref{Eq:F2}) has equal weight for each halo, while the denominator up-weights massive haloes, which yields higher amplitudes for the mean acceleration of the stacked cluster.

In our simulations we find
$F_2$ is quite noisy at $r>3{r_{200}}$. A larger simulation box with a larger halo sample may help to reduce the noise, but mass estimation at such large scales is not our primary concern and we leave its  investigation to a future work.

In summary, within $3{r_{200}}$, the asymmetry of mass distribution for individual haloes leads to a deeper potential along the long axis. Also, halo
centres have non-zero accelerations. When mass weighted, the radial acceleration is boosted compared to its isotropically averaged version. This causes an over-estimation of the mass profile by a few percent at small radii and up to a few 10s of percent at the scales larger than the virial radius. Having a wide range of halo masses in the stacked cluster tends to increase the bias. 
Results from using the centre of mass as the halo centre remain similar except for near the halo centres ($r<0.1r_{200}$), as expected from the offsets between the centre of mass and BCG.

\subsection{Correcting for the total mass bias}

The total correction one needs to apply for halo mass estimation using the conventional Jeans equation is simply the product of $F_1$ and $F_2$. {We show in subplots E \& F of Fig.~\ref{Fig:F1} the inverse of it $1/(F_1F_2)$, which characterises the bias}. The factor $1/(F_1 F_2)$ remains close to unity at the very centre of the halo at $r<0.1 h^{-1} $Mpc when BCGs are chosen as their centres. They deviate more from unity when the centre of mass is used. At slightly larger radii but still within  $r_{200}$, the contribution from mass infall is still not so important, $F_1 F_2$ is dominated by $F_2$. For the conservative case, $1/(F_1F_2)$ is usually larger than unity by a few percent and up to $20\%$ when haloes are binned in narrow mass ranges. The deviation is larger when haloes are selected above a threshold velocity dispersion.

At $r\sim r_{200}$, both $F_1$ and $F_2$ contribute and the mass dependence of $F_1F_2$ is much stronger, 
driven mostly by the mass dependence of $F_1$. The biases are at the $10-20\%$ level for different mass bins. They are greater than 30\% and up to 50\% when haloes are selected above thresholds of velocity dispersion. 

At $2r_{200}<r<3r_{200}$,  $1/(F_1F_2)$ exceeds unity by approximately 30\%, 50\% and 200\% larger than unity for haloes with the mass of $1-2\times 10^{13}h^{-1}$ M$_{\odot}$,  $1-2\times 10^{14}$ and $2-10\times 10^{14}h^{-1}$ M$_{\odot}$ respectively. The product {deviates from unity by approximately a factor of 2} for haloes selected above velocity dispersions thresholds. 

In summary, the bias for halo mass estimate using the velocity dispersion is at the 10s of percent level at the virial radius. Such a bias increases with radius, halo mass and with the range of halo mass in the stacked sample. When haloes are selected using a velocity dispersion threshold, which can include a wide range of halo masses, a factor of 2 overestimation for the true mass is expected.

Finally, we show with dotted lines in Fig.~\ref{Fig:MassProfile} that applying the correction term $F_1F_2$ does 
recover the true mass without any bias. This is evident by the fact that the dotted lines in the top panels lay on top of the dashed lines, and that the dotted lines in the bottom panels do not deviate significantly from unity. The success of $F_1F_2$ in recovering the true mass is numerical justification of our reasoning for the sources of bias in halo mass estimation using the conventional Jeans equation.

\section{Discussion and Concluding Comments}

Much effort is being applied to trying to test, and determine the cosmological
parameters of, the standard $\Lambda$CDM model using the evolution of the
cluster mass function. This is a worthy goal. Here our focus is rather different in that we are primarily interested in trying to test theories of gravity by comparing the mean mass
for a sample of clusters inferred from internal velocity dispersions to
that measured from the mean weak lensing.

A strength of the stacking method is that one can obtain highly equilibrated and
symmetric composites in redshift space, as has already been demonstrated 
by e.g. \citep{Wojtak2011}.  Another strength is that the lensing signal
is relatively strong and so measurements should not be
too susceptible to the kind of systematic
artefacts that plague measurements of the power spectrum of cosmic shear.

The main difficulty in applying this method is in dealing with the
systematic biases in determining the mean dynamical mass from the
velocity dispersion. One such bias stems from the
velocity dispersion anisotropy. Our purpose here has been to explore
and quantify the additional biases. These stem from e.g.\ the heterogeneity
of the clusters combined in the stacks and also their asphericity
and also from the fact that Jeans equation cannot be applied in
its usual form to a composite because the measured velocities
are measured relative to reference particles that are in free-fall
rather than being tied to a non-inertial Newtonian reference frame.

By means of simulations we have been able to confirm the 
presence of the biases expected theoretically and have
attempted to estimate their importance. 
We find that for cluster samples of limited mass range these
biases can be at the $\sim 20 \%$ level at around the virial
radius, and therefore similar to the biases found in simulations
associated with velocity dispersion anisotropies.
Much larger biases are found for `cumulative' samples with a
broad mass range, but it should be emphasised that these would
not be present if, for example, the stacking were performed
using data re-scaled by the individual cluster velocity dispersions.

The results show biases that are relatively well estimated and
show modest and simple trends with the cluster selection criteria. 
It therefore seems promising that application of these methods should be able to
provide fairly precise tests of modifications of gravity such as
`5th forces' operating on the scales of clusters.

There are numerous ways the results here can be extended. It will
be interesting to see how the biases depend on redshift; 
the effect of re-scaling according to velocity dispersion deserves
further study; also interesting is the effect of different choices
for the cluster centre and velocity reference such as taking the
centre of mass of the cluster members. As well as
the prediction for lensing measurements, it is interesting to use
the methods here to sharpen predictions for other observables
such as the gravitational redshift. Another issue is whether
there is any substantial impact from the non-conservation of
galaxies expected as blue/late type galaxies transform into
red/early types as they fall into the cluster and have their gas
stripped.  Finally, there is the prospect of improving the
precision of the estimates by using bigger simulations.

\section*{Data availability}
The halo catalogues from the Millennium simulation used in this paper is publicly available at https://virgodb.dur.ac.uk/. The particle data of the Millennium simulation and results produced in this work are available upon reasonable request to the authors.

\section*{Acknowledgments}
This manuscript was written in 2017 as part of a series of papers on galaxy clusters. Sadly, the lead author, Nick Kaiser, passed away in 2023 before we were able to finish other papers in the series, apart for the first one \citep{Cai2016}. Since this manuscript was essentially completed then and the contents approved by all co-authors, including Nick, we have now decided to submit it posthumously. We have, however, made necessary revision to the manuscript to reflect recent development of the subject. We thank the referee, Wenting Wang, for providing very useful comments.

YC was supported by funding from an STFC Consolidated Grant, the European Research Council under grant number 670193 and the Durham Junior Research Fellowship.
YC acknowledges a grant with the RCUK reference ST/F001166/1. SMC acknowledges the support of the STFC [ST/p000451/1] and ERC [GA 267291]
The simulations and part of data analysis for this paper were performed 
using the DiRAC Data Centric system at Durham University,
operated by the Institute for Computational Cosmology on behalf of the
STFC DiRAC HPC Facility (http://www.dirac.ac.uk). This equipment was funded by
BIS National E-infrastructure capital grant ST/K00042X/1, STFC capital
grant ST/H008519/1, and STFC DiRAC Operations grant ST/K003267/1 and
Durham University. DiRAC is part of the National E-Infrastructure.
Part of the analysis was done on the Geryon cluster at the Centre for Astro-Engineering UC, which
received recent funding from QUIMAL 130008 and Fondequip AIC-57. Access to the simulations used in 
this paper can be obtained from the authors.

\bibliographystyle{mn2e}

\appendix
\section{The simulation set up}

We use the Millennium simulation for this study \citep{Springel2005}. The simulation is run in the concordance $\Lambda$CDM model, 
with $\Omega_{\rm m}=0.25$,  $\Omega_{\rm \Lambda}=0.75$, $h=0.73$, $\sigma_8=0.9$ and $n=1$.
The simulation has $2160^3$ particles in a box of 500 $h^{-1}$Mpc aside. The particle mass is $8.6h^{-1} {\rm M}_{\odot}$, so haloes with the mass of
$M>10^{13}h^{-1}{\rm M}_{\odot}$ have at least $10^4$ particles. The softening length of the force is 5 $h^{-1}$kpc. 
The high resolution of the simulation enable us to probe cluster dynamics deep into the halo centre. 

\subsection{Definitions of haloes and centres}
Dark matter haloes are defined as the most massive sub-haloes in each Friends-of-Friends (FoF) groups in the simulation using the linking length of 0.2 \citep{Davis1985}. The most bound particle within each sub-halo are chosen as the default choice of halo centre \citep{Springel2008}. Halo mass is defined as the mass within the bounding radius of $r_{200}$ within which the mean density is 200 times of the critical density of the universe. We take the averaged quantities including position, velocity and acceleration from particles within a core radius of 30$h^{-1}$kpc for the BCG. 

Note that for the cluster system, we use proper distances and total velocities: $r=ax$, $\dot r=a[Hx +v_{\rm pec}]$, where $x$ is the comoving coordinate, $a$ is the scale factor of the Universe, $H$ is the Hubble constant and $v_{\rm pec}$ is the peculiar velocity. The acceleration is the total one which includes the background contribution from the cosmological constant.

\subsection{Estimating $F_1$}
To evaluate $F_1$ in equation~(\ref{Eq:F1_2}), we need to take the time derivative for the radial momentum density.  We do this by advancing the simulation into the future, from $a=1$ to $a=1.04$. We then difference the momentum density between these two snapshots and divide it by the time interval. To insure that the halo centres correspond to each other between these two epochs, we identified all particles within the 30$h^{-1}$Mpc cores from the halo centres at $a=1$, and use their averaged positions and velocities at $a=1.04$ to define the new centres. 

\subsection{Estimating the isotropic radial acceleration.}

Estimating the numerator in (\ref{Eq:F2}) requires sampling the radial accelerations around each halo isotropically. To do this, we insert spherical grids of massless test particles into the simulation, and use {\sc Gadget} to output the accelerations for them. Starting from the list of halo centres, the test particles are distributed in spherical shells. We have 70 shells distributed logarithmically along the radial direction, ranging from $r=0.01 h^{-1}$Mpc to $35h^{-1}$Mpc, 20 shells for each order of magnitude. For each shell, we
uniformly distribute over $4\pi$ steradians 
768 pixels (particles) using {\sc Healpix} \citep{Gorski2005}. The mean spacing of particles on the sphere is 7.3 deg. Convergence of our results have been tested for the radial and tangential directions by having twice as many bins along each direction. With this setting of test particles and the original dark matter particles in the simulation, we calculate $F_2$ using (\ref{Eq:F2}). Results are shown in Fig.~\ref{Fig:F1}.

\end{document}